\documentclass[final,3p,times]{elsarticle}

\usepackage{amssymb}

\journal{ }

\usepackage{amsmath}
\usepackage{hyperref}

\usepackage[retain-explicit-plus, binary-units=true]{siunitx}

\usepackage{cleveref}

\usepackage{lipsum}

\usepackage{subfig,epsfig,amsfonts}

\usepackage{caption}
\usepackage{lineno}

\usepackage{etoolbox} 
\makeatletter 
\newcommand*\linenomathpatch{\@ifstar{\linenomathpatch@AMS}{\linenomathpatch@}}
\newcommand*\linenomathpatch@[1]{
  \expandafter\pretocmd\csname #1\endcsname {\linenomathWithnumbers}{}{}
  \expandafter\pretocmd\csname #1*\endcsname{\linenomathWithnumbers}{}{}
  \expandafter\apptocmd\csname end#1\endcsname {\endlinenomath}{}{}
  \expandafter\apptocmd\csname end#1*\endcsname{\endlinenomath}{}{}
}
\newcommand*\linenomathpatch@AMS[1]{
  \expandafter\pretocmd\csname #1\endcsname {\linenomathWithnumbersAMS}{}{}
  \expandafter\pretocmd\csname #1*\endcsname{\linenomathWithnumbersAMS}{}{}
  \expandafter\apptocmd\csname end#1\endcsname {\endlinenomath}{}{}
  \expandafter\apptocmd\csname end#1*\endcsname{\endlinenomath}{}{}
}
\let\linenomathWithnumbersAMS\linenomathWithnumbers
\patchcmd\linenomathWithnumbersAMS{\advance\postdisplaypenalty\linenopenalty}{}{}{}
\makeatother 

\linenomathpatch{equation}
\linenomathpatch*{gather}
\linenomathpatch*{multline}
\linenomathpatch*{align}
\linenomathpatch*{alignat}
\linenomathpatch*{flalign}

\begin{document}

\begin{frontmatter}



\title{A method for computing hadron--nucleus cross sections in fully active sampling calorimetric targets}


\author[1]{Johnny Ho\corref{cor1}}

\ead{jh@berkeley.edu}

\address[1]{Department of Physics,
            Harvard University,
            Cambridge, MA 02138, USA}
\address[2]{Enrico Fermi Institute,
            University of Chicago,
            Chicago, IL 60637, USA}

\author[2]{David W.\ Schmitz}

\ead{dwschmitz@uchicago.edu}

\cortext[cor1]{Corresponding author}

\begin{abstract}
Interaction cross section measurements in fixed-target scattering experiments are typically performed by measuring the attenuation of a beam of particles that is incident upon a target slab of material.  A fully active sampling calorimeter, such as the time projection chamber, is a more voluminous target that can be treated as a series of smaller target slabs.  We present a method of computing hadron--nucleus interaction cross sections for data taken with fully active sampling calorimeters that is robust even if the target is comprised of numerous slabs of various thicknesses.
\end{abstract}







\end{frontmatter}


\section{Introduction}
\label{section:introduction}

Fixed-target scattering experiments are a cornerstone of nuclear and particle physics.  Under the guidance of Ernest Rutherford, Hans Geiger and Ernest Marsden performed some of the earliest fixed-target experiments between 1908 and 1913 in which they observed the scattering of an $\alpha$ particle beam upon thin foils of metal~\cite{doi:10.1098/rspa.1908.0067, doi:10.1098/rspa.1909.0054, doi:10.1098/rspa.1910.0038, doi:10.1080/14786440408634197} and provided the experimental evidence that led to the discovery of the atomic nucleus~\cite{doi:10.1080/14786440508637080}.  Historically, fixed-target scattering cross section experiments were typically done by measuring the attenuation or transmission of a hadron ($\pi^\pm$, $n$, $p/\bar{p}$, $K^\pm$) beam through a slab of some nuclear target~\cite{Stroot:1973LftL}.  Over the past 60 years, there have been numerous\footnote{Our compilation of references to these measurements is non-exhaustive.} pion~\cite{CROZON1965567, Wahlig:1968bk, Carter:1968zza, Giacomelli:1969fhx, Binon:1970ye, Carter:1971tj, Wilkin:1973xd, Clough:1974qt, Bertin:1976uh, Carroll:1976hj, Navon:1979wx, Thomas:1979xu, Preedom:1981zz, Ashery:1981tq, Navon:1983xj, Navon:1983zx, Ashery:1986nt, Smith:1989gv, Kohler:1991ht, Clement:1992tn, Kohler:1993me, Jones:1993ps, Kahrimanis:1997if, LADS:1999dyv, Gelderloos:2000ds, Fujii:2001rc, Pavan:2001gu, Lee:2002eq, DUET:2015ybm}, neutron~\cite{Coor:1955zz, Booth_1958, Kriesler:1968zz, Engler:1968ixa, Engler:1968zz, Engler:1970dt, Mischke:1970ph, Jones:1971kba, Schimmerling:1973bb, Longo:1974cw, Jones:1974cy, RamanaMurthy:1975vfu, Biel:1975bs, Barton:1976iv, Stone:1977ih, Stone:1977jh, DeHaven:1978zz, Dehaven:1979vqa, Roberts:1979jf, Winters:1991zz, Boukharouba:2001hk}, proton~\cite{Peelle:1957zz, Meyer:1960zz, Wilkins:1963zza, Pollock:1965zz, Bellettini:1966zz, Bugg:1966zz, Dicello:1967zz, Igo:1967osa, Dicello:1970mx, Greenlees:1971ge, Bray:1971fob, Menet:1971zz, Renberg:1972sg, Renberg:1972jf, Schwaller:1972at, Montague:1973env, Bizzarri:1973sp, McGill:1974zz, Saudinos:1974ef, Carlson:1975zz, Schwaller:1979eu, Ray:1979qk, DeVries:1980zz, Barlett:1980rv, Kalogeropoulos:1980az, Ray:1981nb, Hoffmann:1981uf, Mcgill:1984jq, Nakamura:1984xw, Garreta:1984rs, Balestra:1984wd, Ashford:1985fh, Carlson:1985caw, Balestra:1985kn, Balestra:1985wk, Balestra:1989rb, Carlson:1994fq, Bianconi:2000qr, Klempt:2002ap, Carvalho:2003pza, NA61SHINE:2011dsu, Bianconi:2011zz, NA61SHINE:2020iqu, Aghai-Khozani:2021tvu}, and kaon~\cite{Bugg:1968zz, Denisov:1971jb, Adams:1973elr, Burnstein:1974ax, Damerell:1975kw, Damerell:1977tv, Charles:1977pq, Marlow:1982xt, Dover:1982zh, Siegel:1984br, Krauss:1992ky, Kormanyos:1993zk, Weiss:1994kt, Kormanyos:1995zz, Michael:1991nw} interaction cross section experiments performed at various facilities such as the Los Alamos Meson Physics Facility (LAMPF) in Los Alamos, New Mexico, United States; the Paul Scherrer Institute (PSI) in Villigen, Switzerland; the TRI-University Meson Facility (TRIUMF) in Vancouver, Canada; Lawrence Berkeley National Laboratory (LBNL) in Berkeley, California, United States; Fermi National Accelerator Laboratory (Fermilab) near Batavia, Illinois, United States; Brookhaven National Laboratory (BNL) in Upton, New York, United States; the Princeton-Pennsylvania Accelerator (PPA) in Princeton, New Jersey, United States; the European Laboratory for Particle Physics (CERN) near Geneva, Switzerland; the National Laboratory for High Energy Physics (KEK) in Tsukuba, Ibaraki, Japan; Rutherford Laboratory in Chilton, Oxfordshire, United Kingdom; Institute for High Energy Physics (Serpukhov) in Protvino, Russia; and others.

The liquid argon time projection chamber (LArTPC)~\cite{Willis:1974gi, Nygren:PEP:1974, Chen:1976pp, Nygren:1978rx, Rubbia:1977zz, Chen:1978yh} has become the technology of choice for studying neutrino oscillations over short ($<\SI{1}{\kilo\meter}$) and long ($> \SI{1000}{\kilo\meter}$) baselines in modern neutrino experiments such as MicroBooNE~\cite{MicroBooNE:2016pwy}, SBND~\cite{MicroBooNE:2015bmn, Machado:2019oxb}, ICARUS~\cite{ICARUS:2004wqc}, and DUNE~\cite{DUNE:2020lwj}.  Given that it is a relatively new detector technology, it can benefit from measurements of hadronic cross sections on argon that would reduce systematic uncertainties in the measurements of neutrino oscillations.  Neutrino--argon interactions in the few-GeV energy domain produce large numbers of hadronic particles, and two significant sources of uncertainty come from: (1) the largely unknown dynamics of hadron--nucleon interactions within the target argon nuclei, and (2) the hadron--argon interactions outside the target argon nuclei as the outgoing hadrons propagate through the liquid argon detection medium.  The LArIAT~\cite{Acciarri:2019wgd} and ProtoDUNE~\cite{Abi:2017aow, Abi:2020mwi, DUNE:2021hwx} experiments have taken up the effort of measuring hadronic cross sections in LArTPCs placed in a controlled beam of charged particles at Fermilab and CERN.  The target used in the LArIAT and ProtoDUNE experiments is a volume of liquid argon that is a fully active tracking calorimeter, so the simple method of measuring cross sections on a thin metal foil needs to be extended for a TPC.  This paper explores one such extension with maximum likelihood estimation.

The work presented in this paper is organized into several sections.  We first review a computational method in estimating hadron--nucleus cross sections using maximum likelihood estimation.  Next, we apply the computational method to simulated data of positively charged pions ($\pi^+$) impinged upon targets of liquid argon (LAr), liquid xenon (LXe), and liquid krypton (LKr).  Finally, the application of this method to data taken with a TPC and a few other types of detectors is briefly discussed.

\section{Computation method}
\label{section:computation-method}
A beam of particles traversing through matter will experience an attenuation in its intensity due to the particles scattering out of the beam or undergoing interactions.  Let us consider a beam of particles that is perpendicularly incident upon a target slab of some homogeneous material where $x$ is the depth from the surface of the slab.
The probability for an interaction to occur in the target region $a \leq x \leq b$ is~\cite{Satchler:1980, Cole:2000, Knoll:1979, Duderstadt:1976}
\begin{gather}
  \textrm{Pr}[a \leq X \leq b]
    = \int_a^b \sigma n \; e^{-\sigma n x} \; \mathrm{d}x
\end{gather}
where $\sigma$ is the cross section per nucleon and $n$ is the number density of the target. For a target of uniform density and thickness $\Delta x$, we can choose $a = 0$ and $b = \Delta x$ such that
\begin{gather}
  \textrm{Pr}[0 \leq X \leq \Delta x]
    = \int_0^{\Delta x} \sigma n \; e^{-\sigma n x} \; \mathrm{d}x
    = 1 - e^{-\sigma n \Delta x} \ldotp
\end{gather}
Thus, the interacting and non-interacting probabilities for a target of thickness $\Delta x$ are
\begin{gather}
  \begin{aligned}
    P_\textrm{interacting}^{\Delta x}
      &= 1 - e^{-\sigma n \Delta x}, \\
    P_\textrm{non-interacting}^{\Delta x}
      &= 1 - P_\textrm{interacting}^{\Delta x} = e^{-\sigma n \Delta x} \ldotp
  \end{aligned}
  \label{eq:probabilities}
\end{gather}

Let us suppose that in our laboratory, we have
\begin{enumerate}[(a)]
  \item a particle source that can fire single particles at some fixed kinetic energy, $E$, and
  \item a collection of $N$ targets of various thicknesses.
\end{enumerate}
Let us also suppose that, for whatever reason, we are only able to use each target exactly once.  In our experiment, we want to determine the cross section at a fixed energy, $\sigma(E)$; we set up our experiment in such a way that our single-particle source fires a particle of energy $E$ at a single fixed target of thickness $\Delta x_i$.  After firing a single particle at the target, we must discard the $i\textsuperscript{\,th}$ target and replace it with a new target of different thickness $\Delta x_{\ell}$ ($\ell \neq i$).  Since we have $N$ targets, we are able to perform $N$ independent trials.

Suppose in the outcome of our experiment we have counted $N_\textrm{int}$ interacting particles and $N_\textrm{non-int}$ non-interacting particles\footnote{We note here that by using the naive definition of probability~\cite{Blitzstein:2019}, $P_\textrm{interacting} \equiv \frac{N_{\textrm{int}}}{N_{\textrm{int}} + N_{\textrm{non-int}}}$, we can express Eq.~\ref{eq:probabilities} as
\begin{align*}%
  P_\textrm{interacting} &= 1 - e^{-\sigma n \Delta x} \\%
  e^{-\sigma n \Delta x} &= 1 - P_\textrm{interacting} \\%
  -\sigma n \Delta x &= \ln \left( 1 - P_\textrm{interacting} \right) \\%
  \sigma &= \frac{1}{n \Delta x}  \ln \left( \frac{1}{1 - P_\textrm{interacting}} \right) \\%
         &= \frac{1}{n \Delta x} \ln \left( \frac{N_{\textrm{int}} + N_{\textrm{non-int}}}{N_{\textrm{non-int}}} \right) \ldotp%
\end{align*}%
} (with $N = N_\textrm{int} + N_\textrm{non-int}$).  The likelihood (probability of a dataset given the model parameters) of measuring this outcome is
\begin{gather}
  \begin{aligned}
    \mathcal{L} ( \sigma \; | \; \textrm{data} )
      &= P ( \textrm{data} \; | \; \sigma ) \\
      &= \prod_j^{\makebox[0pt]{\footnotesize $N_\textrm{int}$}} P_\textrm{interacting}^{\Delta x_j}
         \prod_k^{\makebox[0pt]{\footnotesize $N_\textrm{non-int}$}} P_\textrm{non-interacting}^{\Delta x_k} \\
      &= \prod_j^{\makebox[0pt]{\footnotesize $N_\textrm{int}$}} (1 - e^{-\sigma n \Delta x_j})
         \prod_k^{\makebox[0pt]{\footnotesize $N_\textrm{non-int}$}} e^{-\sigma n \Delta x_k}
  \end{aligned}
  \label{eq:likelihood}
\end{gather}
where $\Delta x_j$ is the thickness of the $j\textsuperscript{\,th}$ target (out of $N_\textrm{int}$) in which an interaction occurred, $\Delta x_k$ is the thickness of the $k\textsuperscript{\,th}$ target (out of $N_\textrm{non-int}$) in which there was no interaction, and $\sigma$ is the parameter in our model; we can use maximum likelihood estimation to fit for the parameter $\sigma$.  Taking the natural logarithm of the likelihood (Eq.~\ref{eq:likelihood}) yields
\begin{gather}
  \begin{aligned}
    \ln \mathcal{L}
      &= \sum_j^{\makebox[0pt]{\footnotesize $N_\textrm{int}$}} \ln (1 - e^{-\sigma n \Delta x_j})
       + \sum_k^{\makebox[0pt]{\footnotesize $N_\textrm{non-int}$}} \ln (e^{-\sigma n \Delta x_k}) \\
      &= \sum_j^{\makebox[0pt]{\footnotesize $N_\textrm{int}$}} \ln (1 - e^{-\sigma n \Delta x_j})
       + \sum_k^{\makebox[0pt]{\footnotesize $N_\textrm{non-int}$}} (-\sigma n \Delta x_k)
      \ldotp
  \end{aligned}
  \label{eq:loglikelihood}
\end{gather}
To maximize the likelihood, we take the partial derivative of Eq.~\ref{eq:loglikelihood} with respect to the fit parameter $\sigma$ and set it to zero, $\frac{\partial}{\partial \sigma} \ln \mathcal{L} = 0$,
\begin{gather}
  \begin{aligned}
    \frac{\partial}{\partial \sigma} \ln \mathcal{L}
      &= \sum_j^{\makebox[0pt]{\footnotesize $N_\textrm{int}$}}
         \frac{\partial}{\partial \sigma} \ln (1 - e^{-\sigma n \Delta x_j})
       + \sum_k^{\makebox[0pt]{\footnotesize $N_\textrm{non-int}$}}
         \; \frac{\partial}{\partial \sigma} (-\sigma n \Delta x_k) \\
      &= \sum_j^{\makebox[0pt]{\footnotesize $N_\textrm{int}$}}
         \frac{(n \Delta x_j) \, e^{-\sigma n \Delta x_j}}{1 - e^{-\sigma n \Delta x_j}}
       - \sum_k^{\makebox[0pt]{\footnotesize $N_\textrm{non-int}$}}
         n \Delta x_k \\
      &= \sum_j^{\makebox[0pt]{\footnotesize $N_\textrm{int}$}}
         \frac{n \Delta x_j}{e^{\sigma n \Delta x_j} - 1}
       - \sum_k^{\makebox[0pt]{\footnotesize $N_\textrm{non-int}$}}
         n \Delta x_k = 0 \ldotp
      \label{eq:derivative-loglikelihood}
  \end{aligned}
\end{gather}
Computationally, we can run a minimizer program to minimize the negative log-likelihood (which is equivalent to maximizing the likelihood), i.e., minimizing the negative of Eq.~\ref{eq:loglikelihood}
\begin{gather}
  - \ln \mathcal{L}
    = - \sum_j^{\makebox[0pt]{\footnotesize $N_\textrm{int}$}} \ln (1 - e^{-\sigma n \Delta x_j})
    + \sum_k^{\makebox[0pt]{\footnotesize $N_\textrm{non-int}$}} \sigma n \Delta x_k
  \label{eq:negative-loglikelihood}
\end{gather}
while varying the fit parameter $\sigma$.  We will refer to this as the \emph{maximum likelihood estimation} (MLE) method.

\subsection{Approximations}
\label{subsection:approximations}

In this subsection, we discuss two different approximations of the log-likelihood function (Eq.~\ref{eq:loglikelihood}).  For the first approximation, let us assume that $\Delta x_j = \Delta x_k = \langle \Delta x \rangle$ for all $j,k$, i.e., all targets have the same exact thickness.  Eq.~\ref{eq:derivative-loglikelihood} then becomes
\begin{subequations}
  \begin{align}
    \frac{\partial}{\partial \sigma} \ln \mathcal{L}
      &= \sum_j^{\makebox[0pt]{\footnotesize $N_\textrm{int}$}}
         \frac{n \Delta x_j}{e^{\sigma n \Delta x_j} - 1}
       - \sum_k^{\makebox[0pt]{\footnotesize $N_\textrm{non-int}$}}
         n \Delta x_k
       = 0
      \tag{\ref{eq:derivative-loglikelihood}} \\
      &= \sum_j^{\makebox[0pt]{\footnotesize $N_\textrm{int}$}}
         \frac{n \langle \Delta x \rangle}{e^{\sigma n \langle \Delta x \rangle} - 1}
       - \sum_k^{\makebox[0pt]{\footnotesize $N_\textrm{non-int}$}}
         n \langle \Delta x \rangle
      \\
      &=
         n \langle \Delta x \rangle
         \left(
         \frac{1}{e^{\sigma n \langle \Delta x \rangle} - 1}
         \sum_j^{N_\textrm{int}}
       - \!\!\sum_k^{N_\textrm{non-int}}
         \right)
      \\
      &=
         n \langle \Delta x \rangle
         \left(
         \frac{1}{e^{\sigma n \langle \Delta x \rangle} - 1}
         N_\textrm{int}
       - N_\textrm{non-int}
         \right)
      = 0
  \end{align}
\end{subequations}
which can then be solved to yield
\begin{gather}
    \frac{N_\textrm{int}}{N_\textrm{int}+N_\textrm{non-int}}
    = 1 - e^{-\sigma n \langle \Delta x \rangle}
    \ldotp
  \label{eq:width-assumption}
\end{gather}
Let us now further assume that $\sigma n \langle \Delta x \rangle$ is small enough such that the higher-order terms, $\mathcal{O}[(\sigma n \langle \Delta x \rangle)^2]$, are negligible.  Under this assumption, we can make the following approximation in Eq.~\ref{eq:width-assumption}
\begin{subequations}
  \begin{align}
    \frac{N_\textrm{int}}{N_\textrm{int}+N_\textrm{non-int}}
      &=
      1 - e^{-\sigma n \langle \Delta x \rangle}
      \tag{\ref{eq:width-assumption}} \\
    \phantom{\frac{N_\textrm{int}}{N_\textrm{int}+N_\textrm{non-int}}}
      &\approx
        1 - \left\{ 1 - \sigma n \langle \Delta x \rangle + \mathcal{O}\left[(- \sigma n \langle \Delta x \rangle)^2\right] \right\} \\
    \phantom{\frac{N_\textrm{int}}{N_\textrm{int}+N_\textrm{non-int}}}
      &\approx
        \sigma n \langle \Delta x \rangle
      \ldotp
  \end{align}
\end{subequations}
This leaves us with
\begin{gather}
  \sigma \approx
    \frac{1}{n \langle \Delta x \rangle} \frac{N_\textrm{int}}{N_\textrm{int}+N_\textrm{non-int}}
  \label{eq:simple-ratio}
\end{gather}
which we will call the \emph{simple ratio} (SR) method.  This is the method that is used, for example, in the LArIAT experiment~\cite{LArIAT:2021yix}.

For the second approximation, we do not assume that $\Delta x_j = \Delta x_k = \langle \Delta x \rangle$ for all $j,k$, but let us still assume that $\sigma n \langle \Delta x \rangle$ is small enough such that the higher-order terms, $\mathcal{O}[(\sigma n \langle \Delta x \rangle)^2]$, are negligible.  Under this assumption, we can make the following approximation in Eq.~\ref{eq:derivative-loglikelihood}
\begin{subequations}
  \begin{align}
    \frac{\partial}{\partial \sigma} \ln \mathcal{L}
      &= \sum_j^{\makebox[0pt]{\footnotesize $N_\textrm{int}$}}
         \frac{n \Delta x_j}{e^{\sigma n \Delta x_j} - 1}
       - \sum_k^{\makebox[0pt]{\footnotesize $N_\textrm{non-int}$}}
         n \Delta x_k
       = 0
      \tag{\ref{eq:derivative-loglikelihood}} \\
      &\approx
         \sum_j^{\makebox[0pt]{\footnotesize $N_\textrm{int}$}}
         \frac{n \Delta x_j}{\left\{ 1 + \sigma n \Delta x_j + \mathcal{O}\left[(\sigma n \Delta x_j)^2\right] \right\} - 1}
       - \sum_k^{\makebox[0pt]{\footnotesize $N_\textrm{non-int}$}}
         n \Delta x_k \\
      &\approx
         \sum_j^{\makebox[0pt]{\footnotesize $N_\textrm{int}$}}
         \frac{n \Delta x_j}{\sigma n \Delta x_j}
       - \sum_k^{\makebox[0pt]{\footnotesize $N_\textrm{non-int}$}}
         n \Delta x_k \\
      &\approx
         \sum_j^{\makebox[0pt]{\footnotesize $N_\textrm{int}$}}
         \frac{1}{\sigma}
       - \sum_k^{\makebox[0pt]{\footnotesize $N_\textrm{non-int}$}}
         n \Delta x_k \\
      &\approx
         \frac{1}{\sigma}
         \sum_j^{\makebox[0pt]{\footnotesize $N_\textrm{int}$}}
       \; - \;
         \sum_k^{\makebox[0pt]{\footnotesize $N_\textrm{non-int}$}}
         n \Delta x_k \\
      &\approx
         \frac{1}{\sigma} N_\textrm{int}
       - \sum_k^{\makebox[0pt]{\footnotesize $N_\textrm{non-int}$}}
         n \Delta x_k \approx 0 \ldotp
  \end{align}
\end{subequations}
This leaves us with
\begin{gather}
  \sigma
  \approx
  \frac{1}{n}
  \frac
  {
    N_\textrm{int}
  }
  {
    \displaystyle
    \sum_k^{N_\textrm{non-int}} \!\! \Delta x_k
  }
  =
  \frac{1}{n \langle \Delta x \rangle}
  \frac
  {
    N_{\textrm{int}}
  }
  {
    \displaystyle
    \sum_k^{N_{\textrm{non-int}}} \!\! \frac{\Delta x_k}{\langle \Delta x \rangle}
  }
  \label{eq:weighted-ratio}
\end{gather}
which we will call the \emph{weighted ratio} (WR) method.

In this paper, we focus on computing the hadronic cross section as a function of kinetic energy, $\sigma(E)$.  One of the underlying assumptions in this method is that $\sigma$ should not vary too much within the energy range of interest, i.e., $\Delta \sigma$ should be small for some $\Delta E$.  In other words, the width of the energy bins should be chosen such that the expected cross section is relatively constant within the energy bins.

\section{Simulations}
\label{section:simulations}

To test the computation method and its approximations described in Section~\ref{section:computation-method}, we used \textsc{Geant}4~\cite{Agostinelli:2002hh, Allison:2006ve, Allison:2016lfl} to simulate positive pions ($\pi^+$) of various energies impinging upon nuclear targets of various thicknesses; the nuclear targets used in the simulations are liquid argon (LAr), liquid krypton (LKr), and liquid xenon (LXe).

\begin{figure}[tp]%
  \centering%
  \subfloat[Target thickness of \SI{1}{\mm}; SR method]%
  {%
    \includegraphics[width=0.25\textheight]{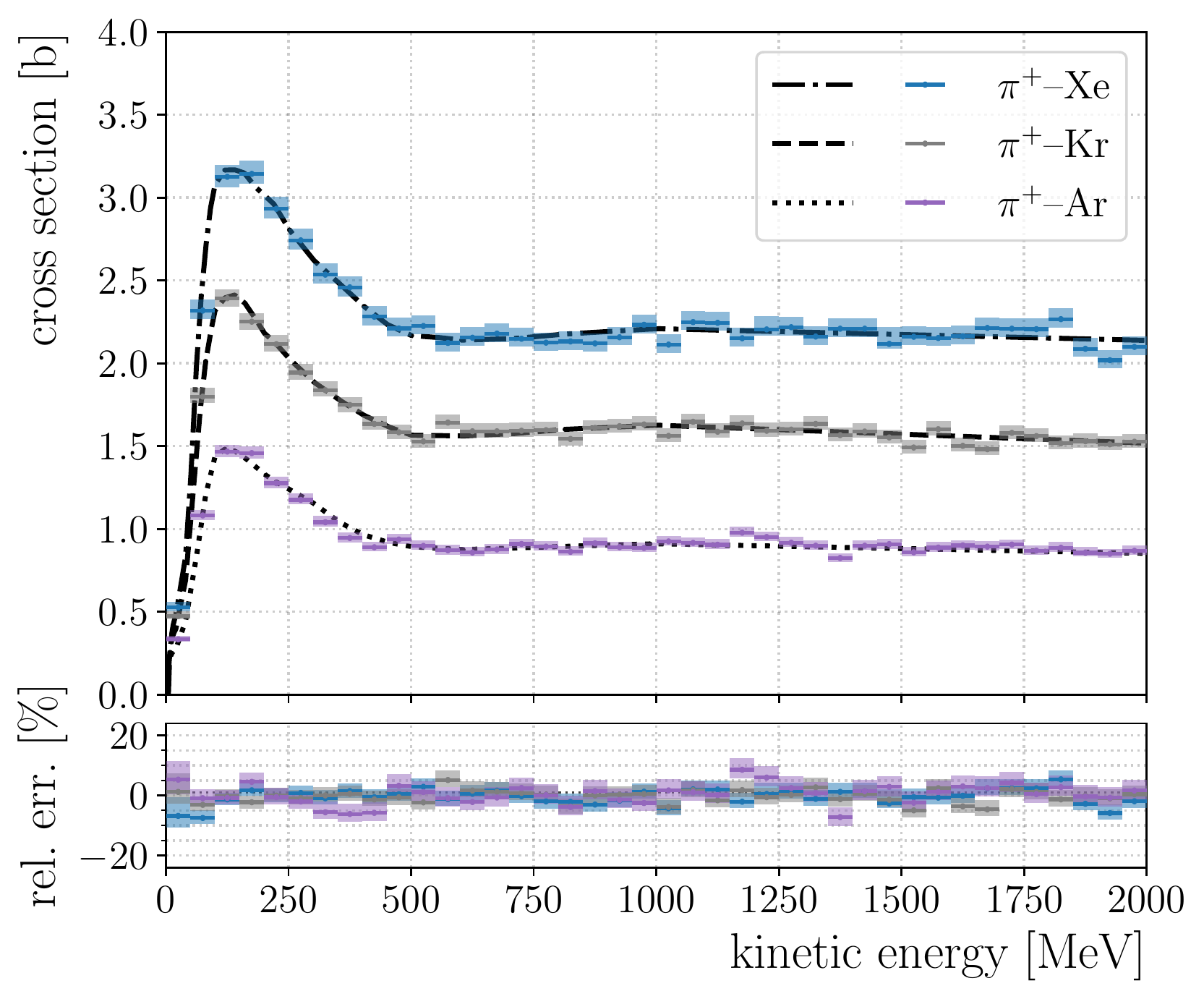}%
  }%
  \hfill%
  \subfloat[Target thickness of \SI{1}{\mm}; WR method]%
  {%
    \includegraphics[width=0.25\textheight]{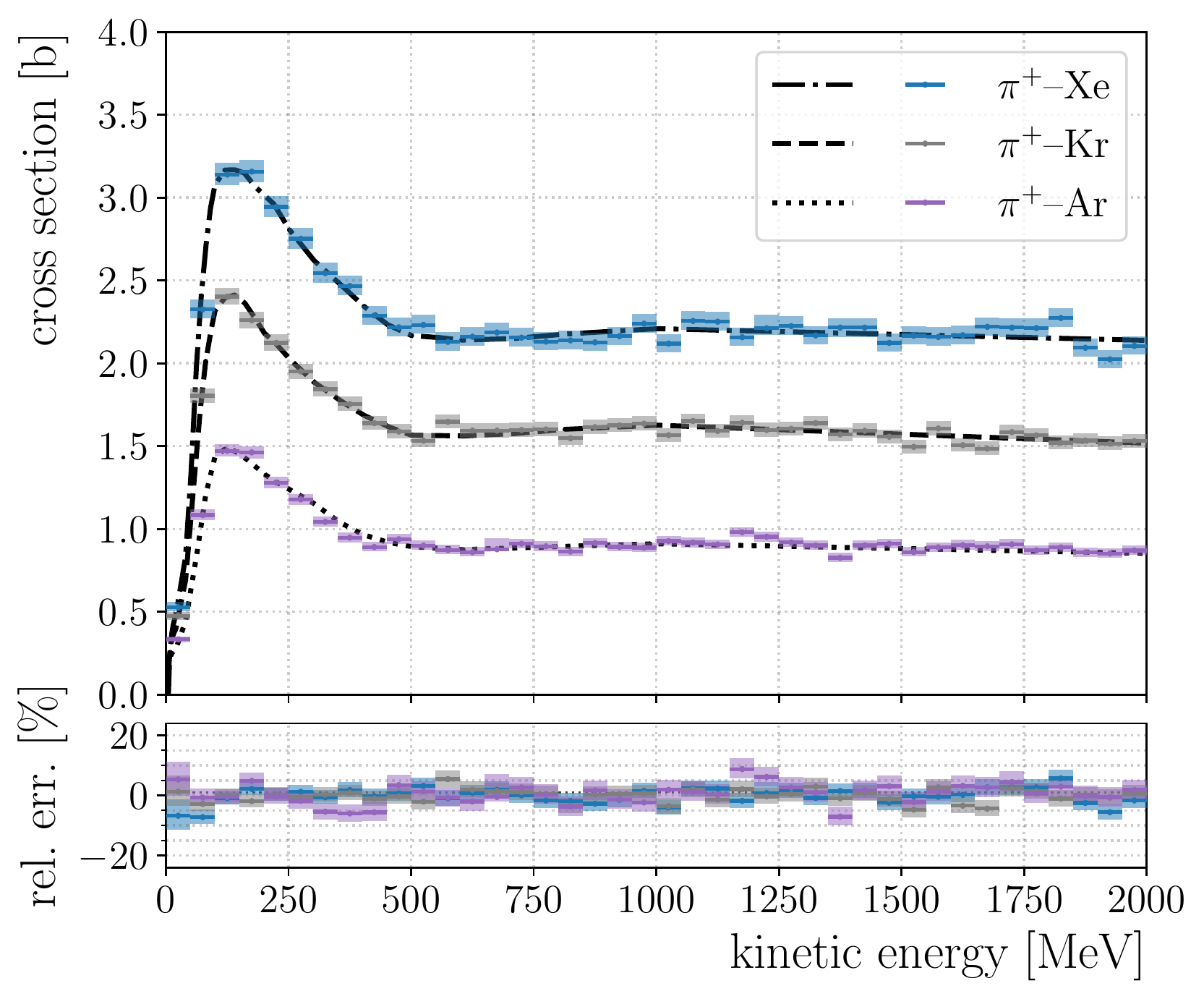}%
  }%
  \hfill%
  \subfloat[Target thickness of \SI{1}{\mm}; MLE method]%
  {%
    \includegraphics[width=0.25\textheight]{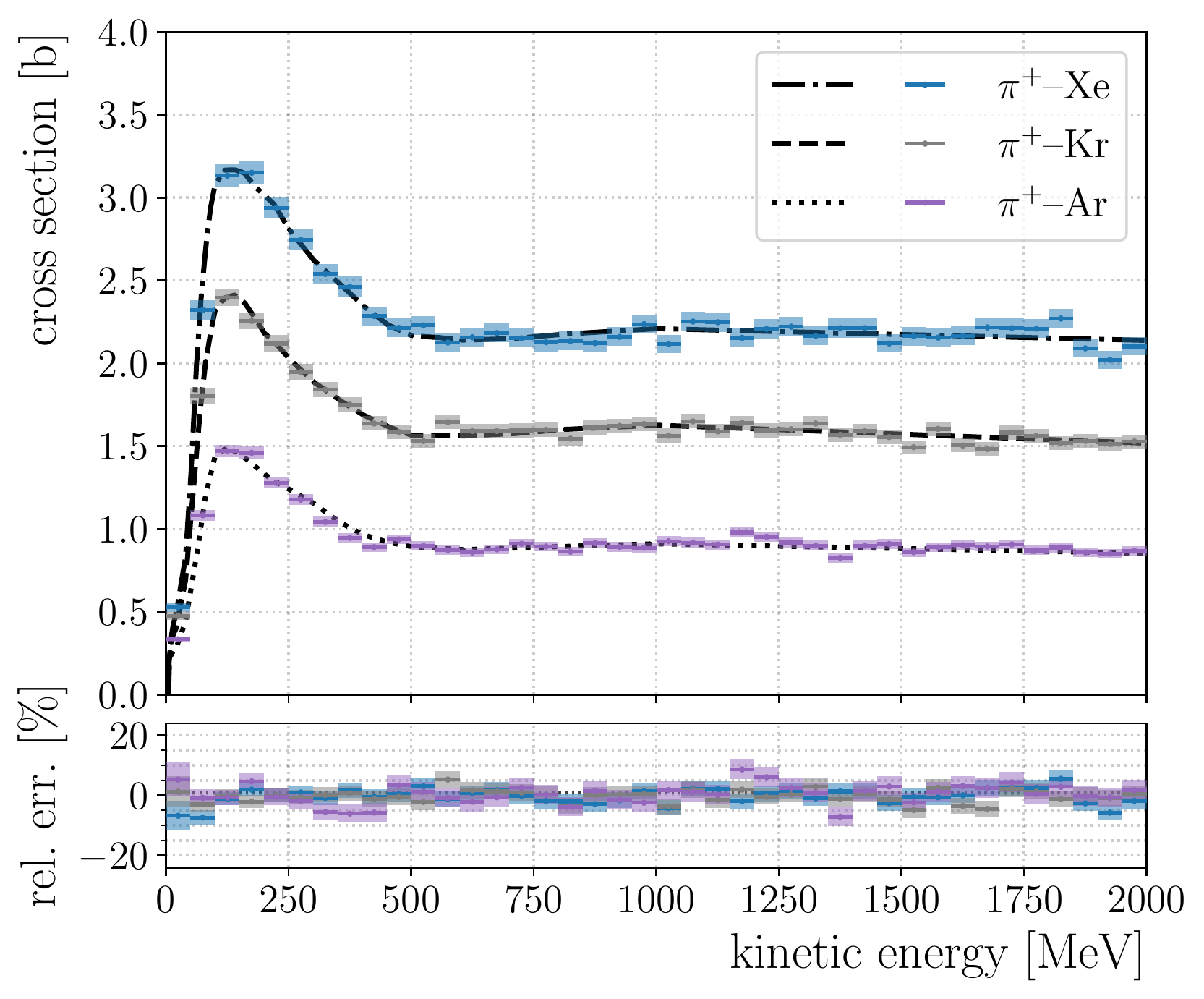}%
  }%
  \\[0.5\baselineskip]
  \subfloat[Target thickness of \SI{20}{\mm}; SR method]%
  {%
    \includegraphics[width=0.25\textheight]{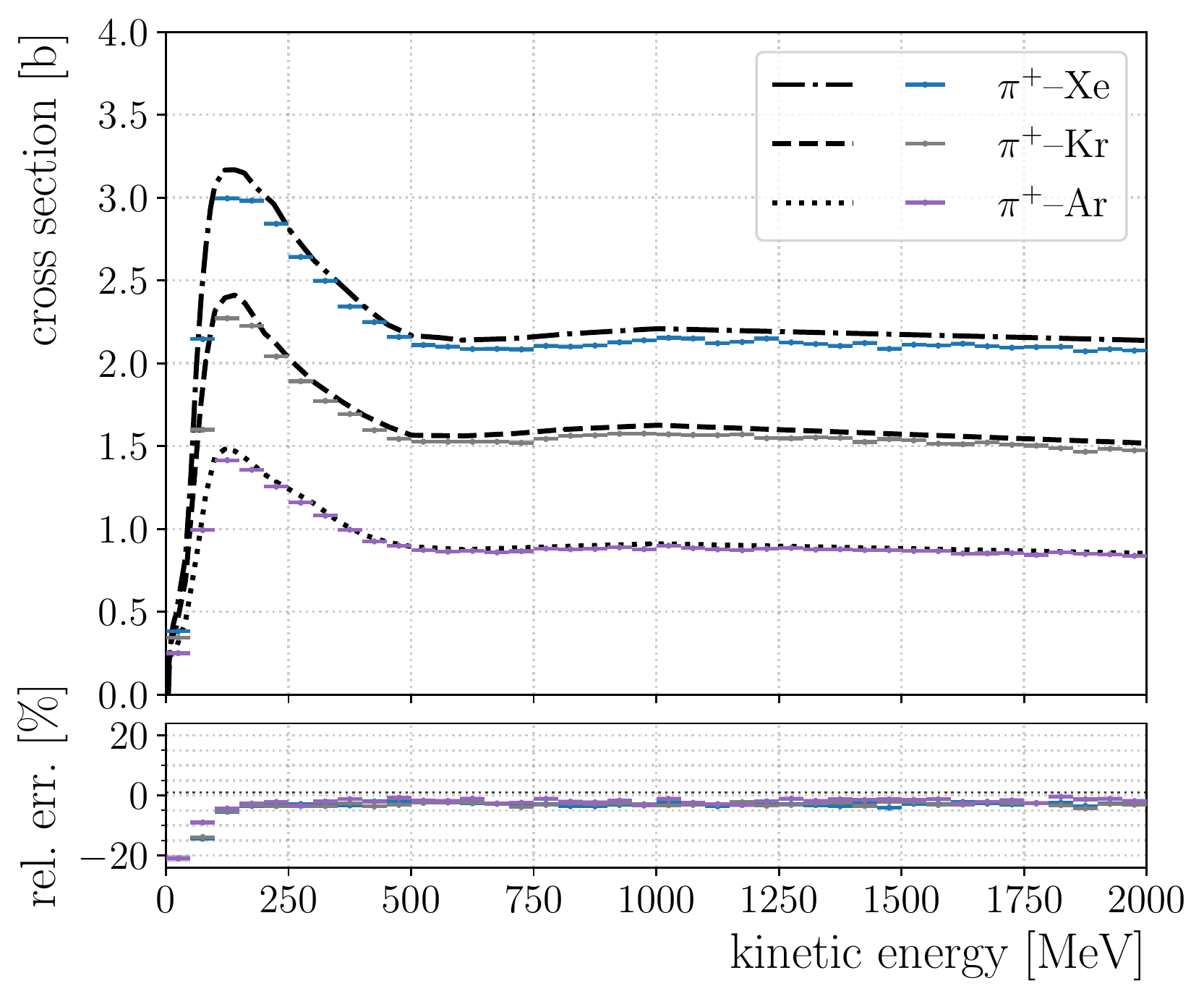}%
  }%
  \hfill%
  \subfloat[Target thickness of \SI{20}{\mm}; WR method]%
  {%
    \includegraphics[width=0.25\textheight]{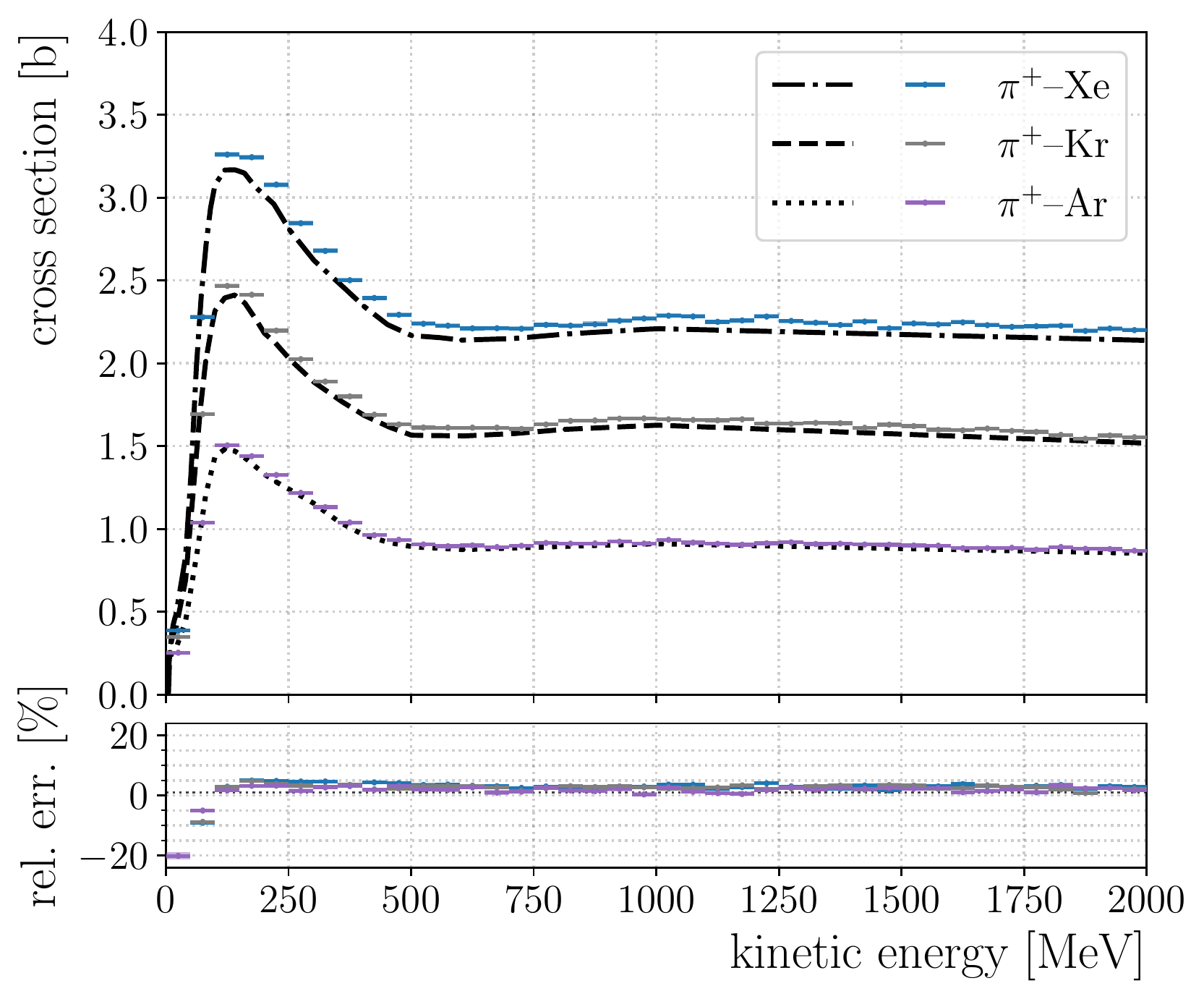}%
  }%
  \hfill%
  \subfloat[Target thickness of \SI{20}{\mm}; MLE method]%
  {%
    \includegraphics[width=0.25\textheight]{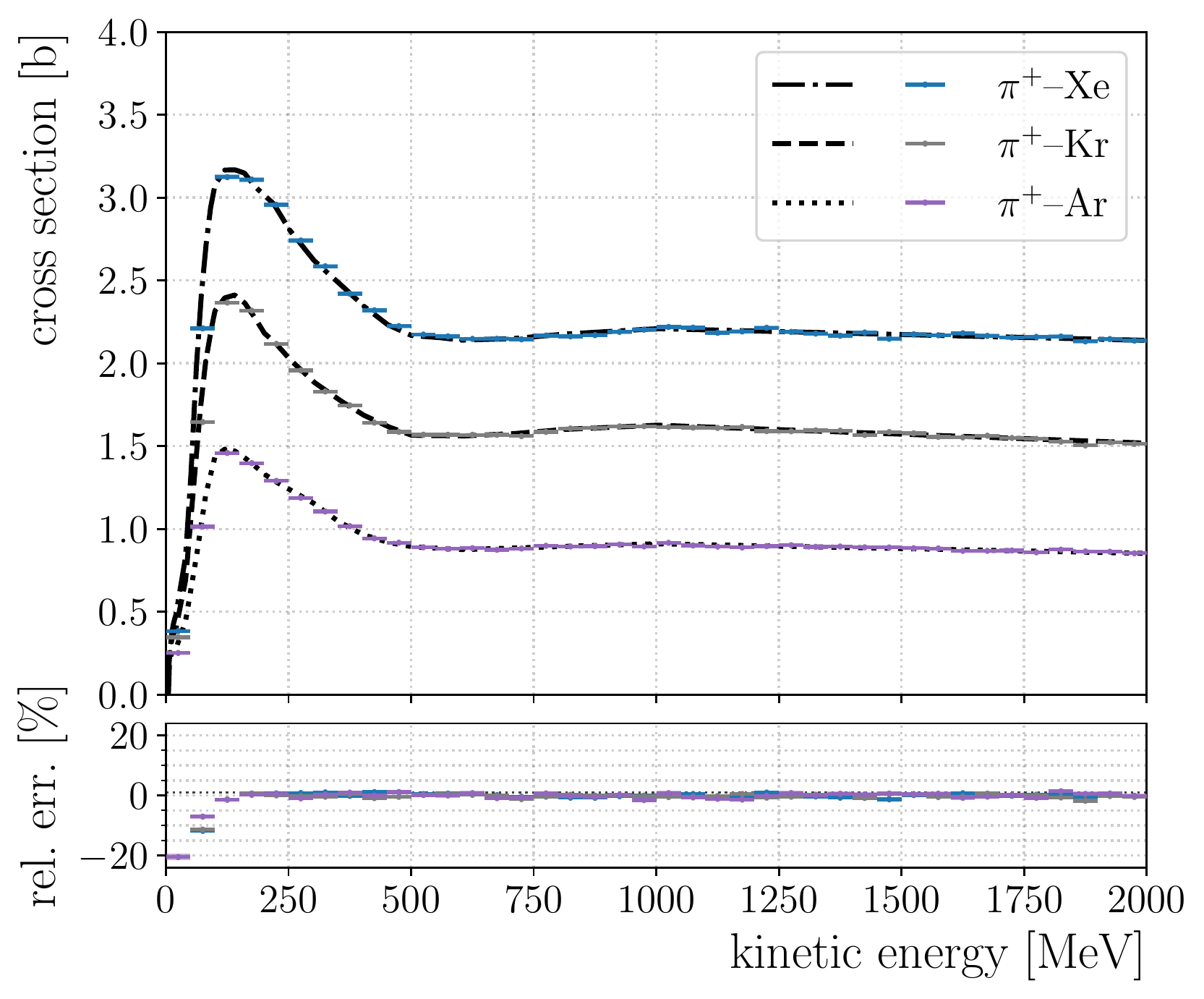}%
  }%
  \\[0.5\baselineskip]
  \subfloat[Target thickness of \SI{80}{\mm}; SR method]%
  {%
    \includegraphics[width=0.25\textheight]{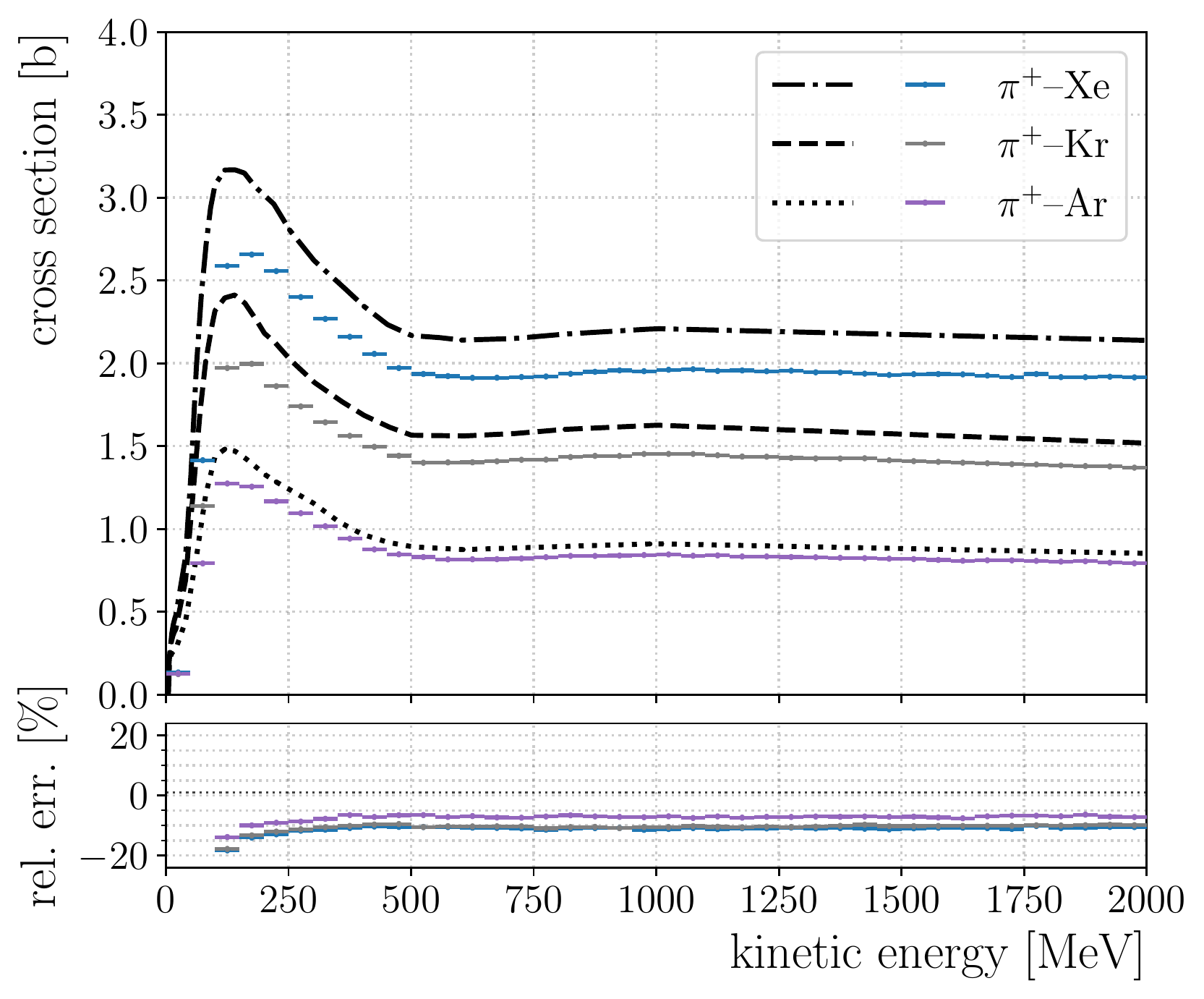}%
  }%
  \hfill%
  \subfloat[Target thickness of \SI{80}{\mm}; WR method]%
  {%
    \includegraphics[width=0.25\textheight]{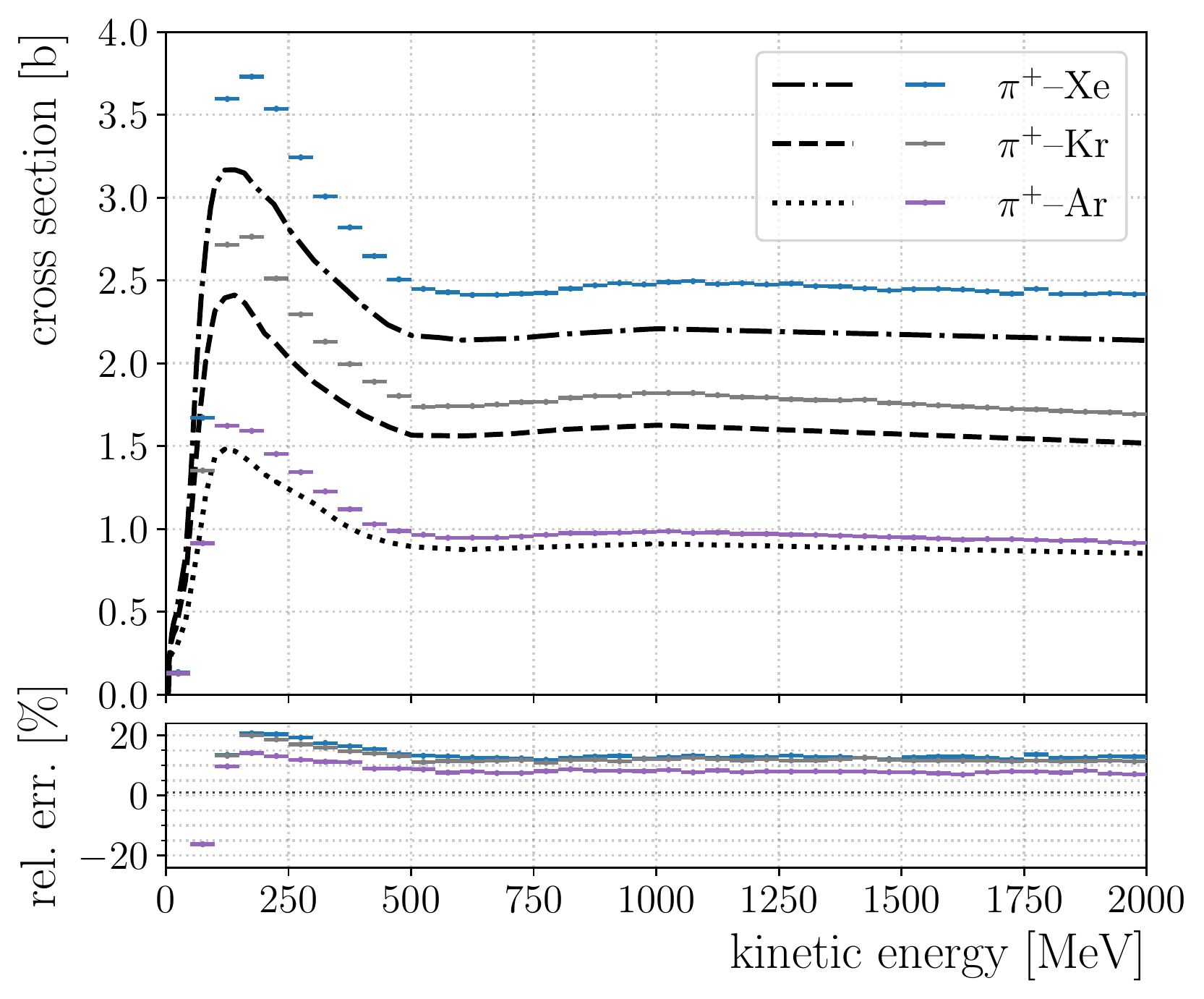}%
  }%
  \hfill%
  \subfloat[Target thickness of \SI{80}{\mm}; MLE method]%
  {%
    \includegraphics[width=0.25\textheight]{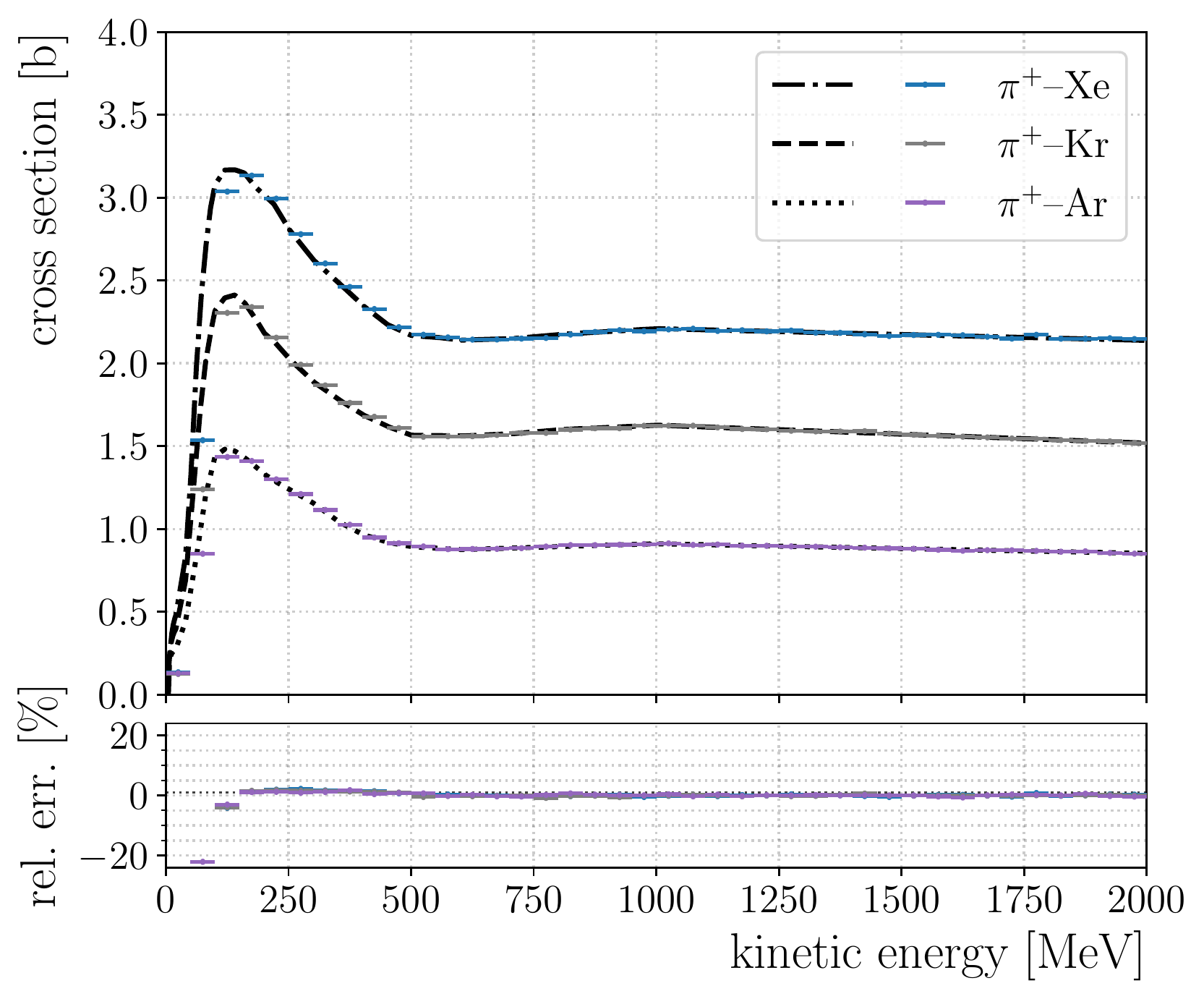}%
  }%
  \caption{Comparison of different methods used to compute the cross sections of $\pi^+$ on LAr, LKr, and LXe targets of different thicknesses.  The underlying cross section models from \textsc{Geant}4 are shown in black dotted, dashed, and dash-dotted lines for $\pi^+$\hbox{--}$\textrm{Ar}$, $\pi^+$\hbox{--}$\textrm{Kr}$, and $\pi^+$\hbox{--}$\textrm{Xe}$, respectively.}%
  \label{fig:sr-wr-mle-grid}%
\end{figure}

\begin{figure}[tp]%
  \centering%
  \subfloat[LAr target]%
  {%
    \includegraphics[width=0.25\textheight]{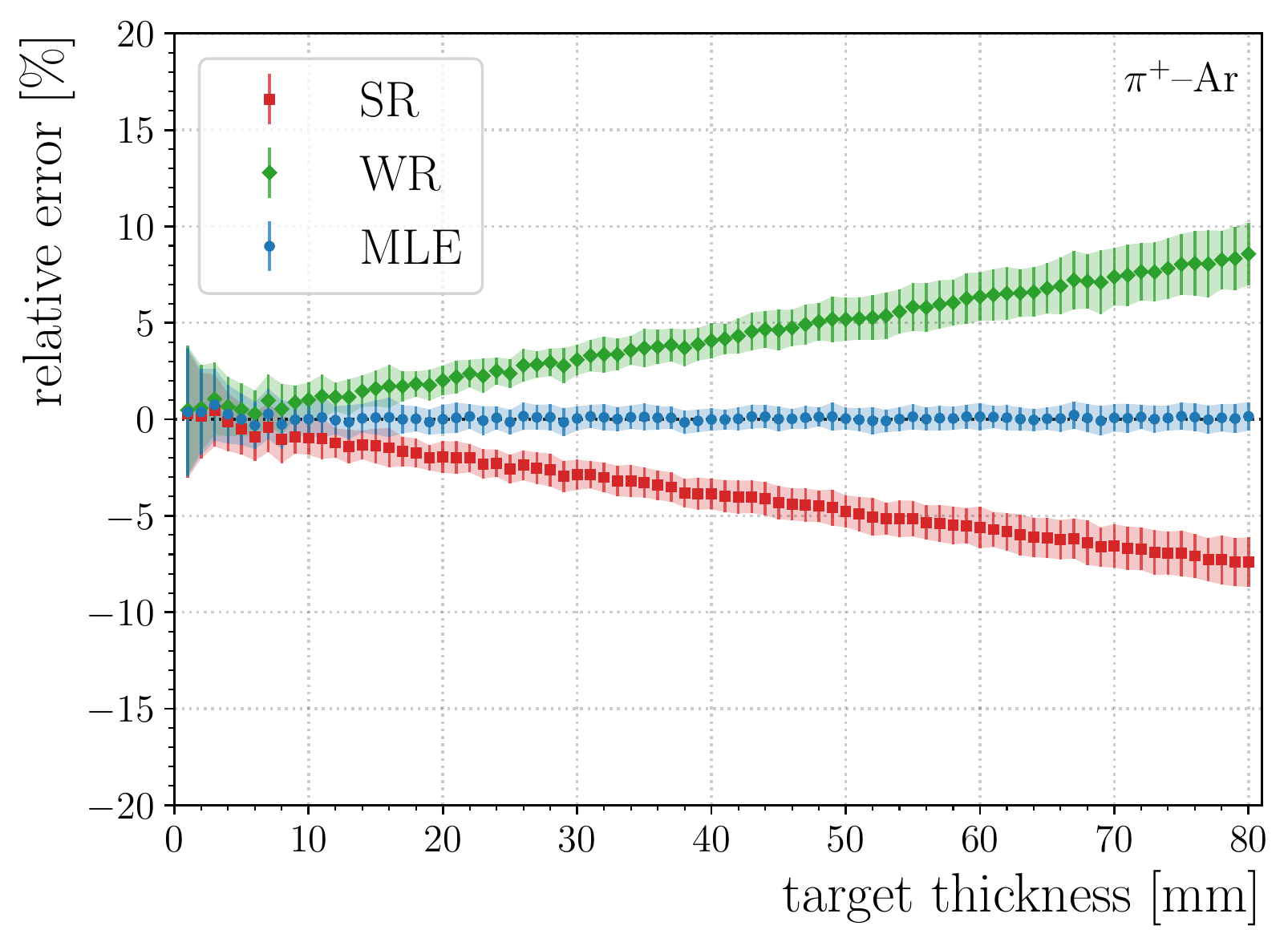}%
  }%
  \hfill%
  \subfloat[LKr target]%
  {%
    \includegraphics[width=0.25\textheight]{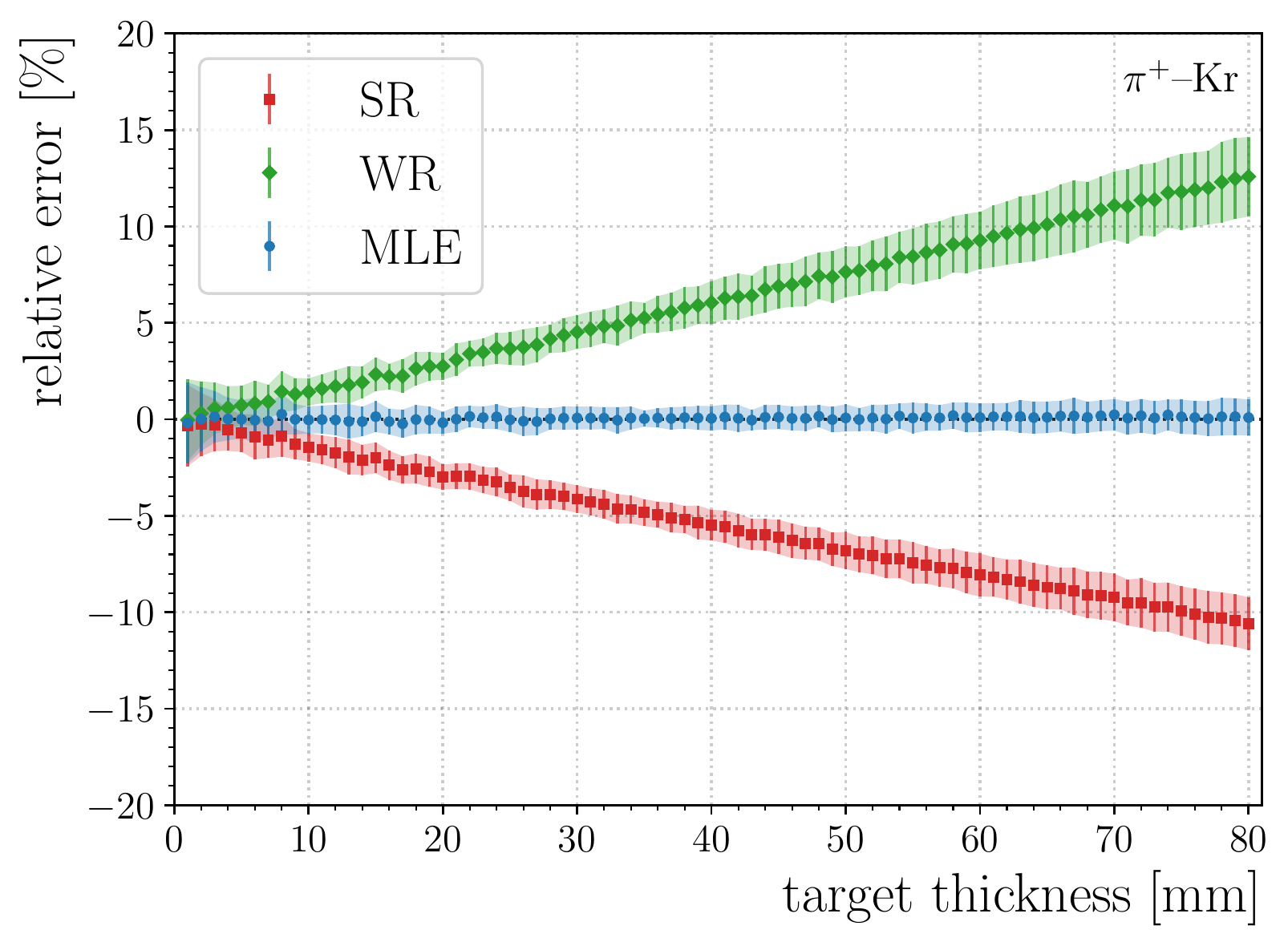}%
  }%
  \hfill%
  \subfloat[LXe target]%
  {%
    \includegraphics[width=0.25\textheight]{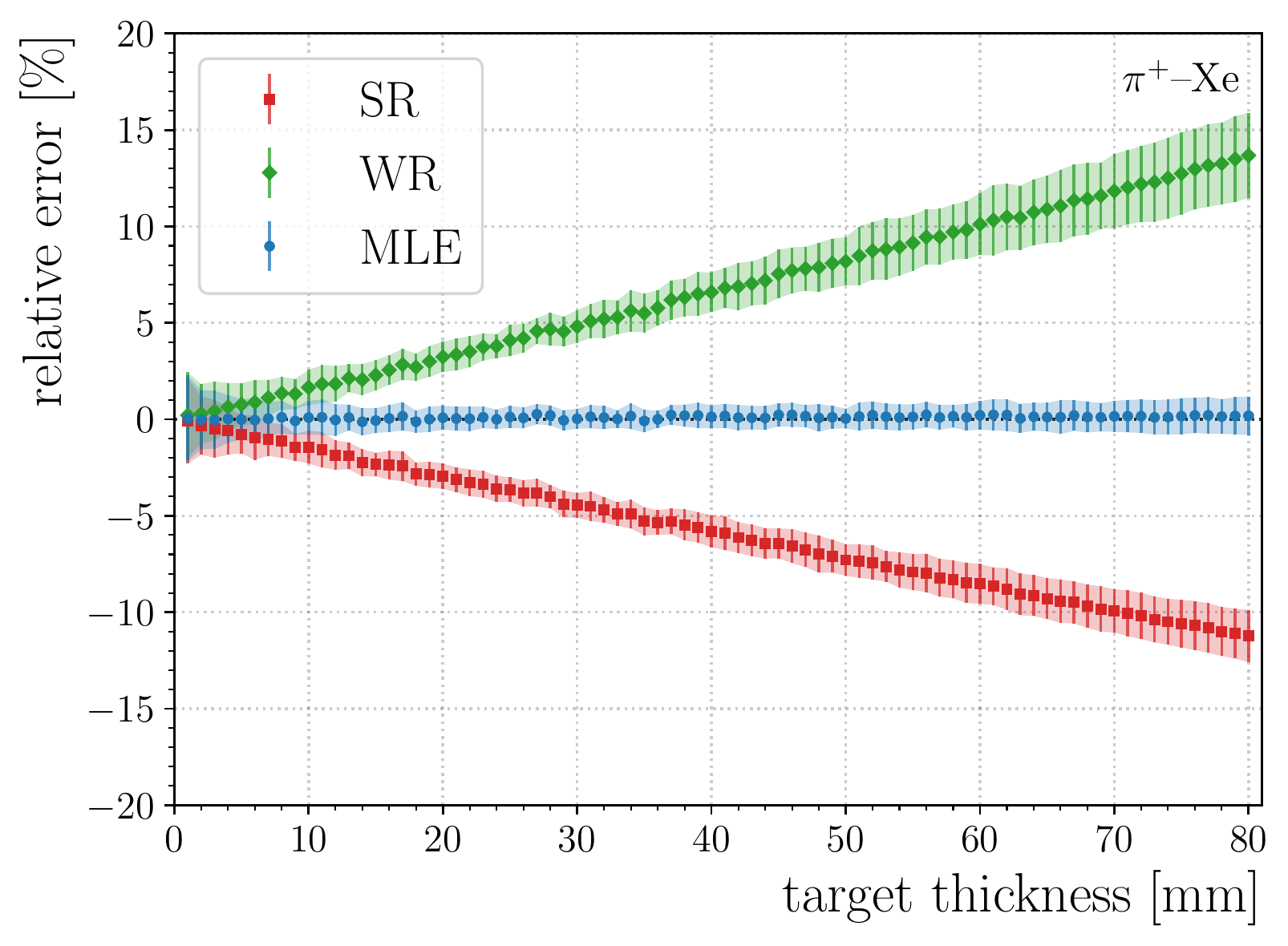}%
  }%
  \caption{Relative error of estimated $\pi^+$\hbox{--}nucleus cross section
           values as a function of target thickness for LAr, LKr, and LXe
           targets.
           The cross section values estimated using the simple ratio (SR) and
           weighted ratio (WR) methods diverge from the true cross section
           values as the target thickness increases;  the cross section values
           estimated using the maximum likelihood estimation (MLE) method
           remain consistent with the true cross section values as the target
           thickness increases.}%
  \label{fig:relative-error-vs-target-thickness}%
\end{figure}

We studied the dependency of Eqs.~\ref{eq:loglikelihood}, \ref{eq:simple-ratio}, and \ref{eq:weighted-ratio} on target thickness by simulating a single-particle gun that fired $\pi^+$ particles at fixed targets of thicknesses from \SIrange{1}{80}{\mm} in steps of \SI{1}{\mm}.  For each target thickness, the single-particle gun fired $\pi^+$ particles with a uniform incident kinetic energy distribution between \SIrange[range-phrase=\text{ and }]{0}{2000}{\MeV} in the direction normal to the surface of the target.  We estimated the total hadronic $\pi^+$\hbox{--}nucleus cross section in \SI{50}{MeV}-wide kinetic energy bins from \SIrange{0}{2000}{\MeV} and compared the results to the true hadronic cross section in \textsc{Geant}4 (this is essentially a closure test).  Fig.~\ref{fig:sr-wr-mle-grid} shows estimated hadronic cross sections of $\pi^+$\hbox{--}Ar, $\pi^+$\hbox{--}Kr, and $\pi^+$\hbox{--}Xe for target thicknesses of \SI{1}{\mm}, \SI{20}{\mm}, and \SI{80}{\mm}.
The true $\left\vert \frac{\mathrm{d}\sigma}{\mathrm{d}E} \right\vert$ values are relatively high for $E < \SI{100}{\MeV}$, resulting in the cross section values being underestimated in the first two kinetic energy bins for all three methods (this effect can be reduced by choosing finer energy bins).  We find that this effect becomes larger as the target thickness increases; in the MLE method, $\sigma$ tends to be underestimated in regions where $\frac{\mathrm{d}\sigma}{\mathrm{d}E} > 0$ and overestimated in regions where $\frac{\mathrm{d}\sigma}{\mathrm{d}E} < 0$.
Fig.~\ref{fig:relative-error-vs-target-thickness} shows the mean relative error of the estimated cross section values (in energy bins where $E > \SI{100}{\MeV}$) as a function of target thickness for LAr, LKr, and LXe targets.  The cross section values estimated using the SR (Eq.~\ref{eq:simple-ratio}) and WR (Eq.~\ref{eq:weighted-ratio}) methods diverge from the true cross section values as the target thickness increases; the SR method tends to underestimate the cross section values whereas the WR method tends to overestimate the cross section values.  The cross section values estimated using the MLE method (Eq.~\ref{eq:loglikelihood}) remain consistent with the true cross section values as the target thickness increases.

\begin{figure}[tp]%
  \centering%
  \includegraphics[width=1.0\textwidth]{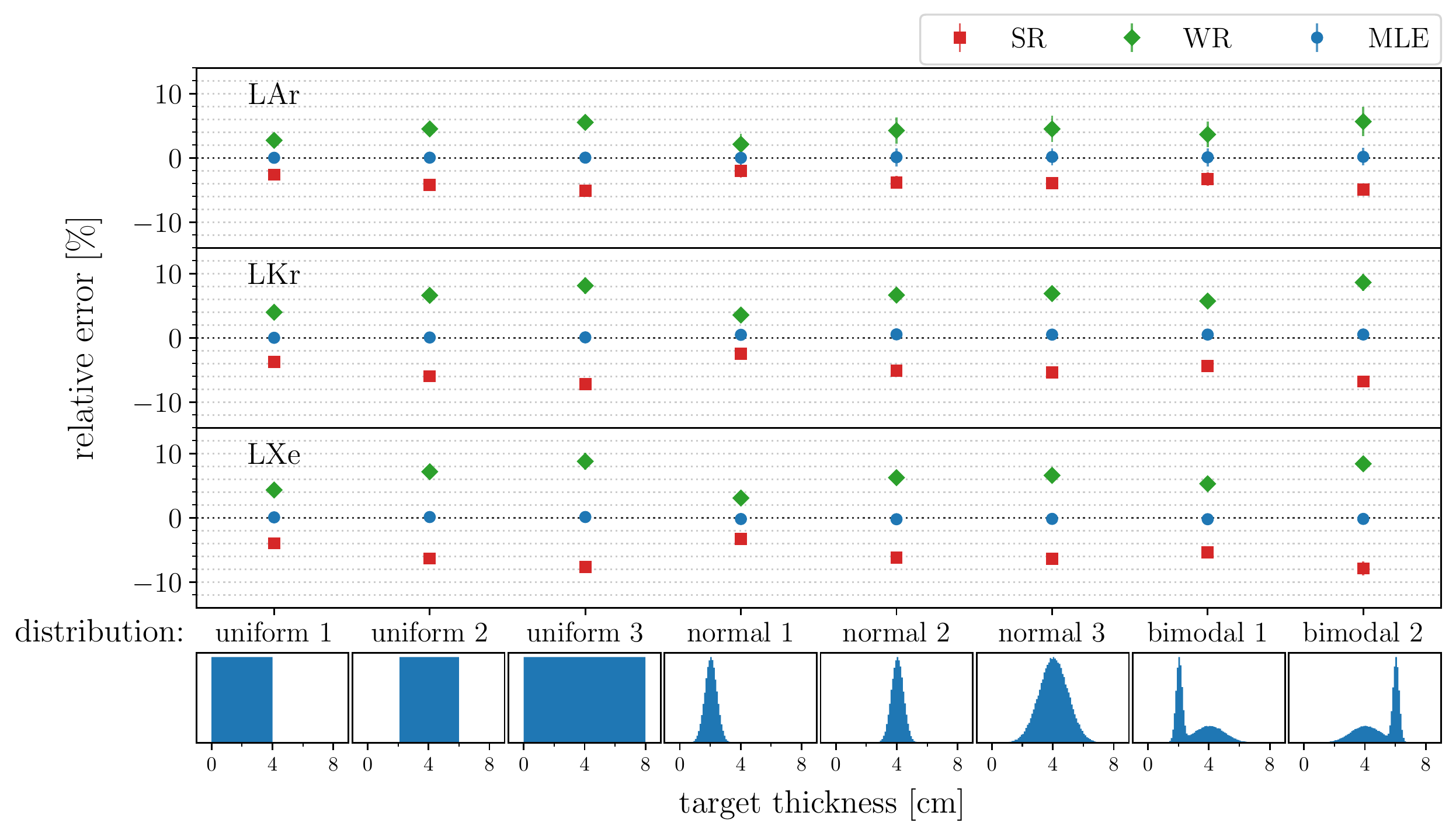}%
  \caption{Relative errors of estimated $\pi^+$\hbox{--}nucleus cross section values using various distributions of target thickness for LAr, LKr, and LXe targets.  The deviation from the true cross section values becomes larger as the mean target thickness of the tested distributions is increased for the SR and WR methods; the mean of a distribution has a larger impact on the relative error than the variance for the SR and WR methods.  The cross section values estimated using the MLE method remain consistent with the true cross section values in all the tested distributions of target thickness.}%
  \label{fig:relative-error-vs-target-thickness-dist}%
\end{figure}

Additionally, we investigated the effects of using uniform, normal, and bimodal distributions of target thickness as shown in Fig.~\ref{fig:relative-error-vs-target-thickness-dist}.  In the SR method, the mean of each distribution was used for $\langle \Delta x \rangle$ in the denominator of Eq.~\ref{eq:simple-ratio}.  We find that the cross section values are underestimated with the SR method and overestimated with the WR method; the deviation from the true cross section values becomes larger as the mean target thickness of the tested distributions is increased.  The cross section values estimated using the MLE method remain consistent with the true cross section values in all the tested distributions of target thickness.

\section{Applications}
\label{section:applications}

\begin{figure}[tp]%
  \centering%
  \includegraphics[width=1.0\textwidth]{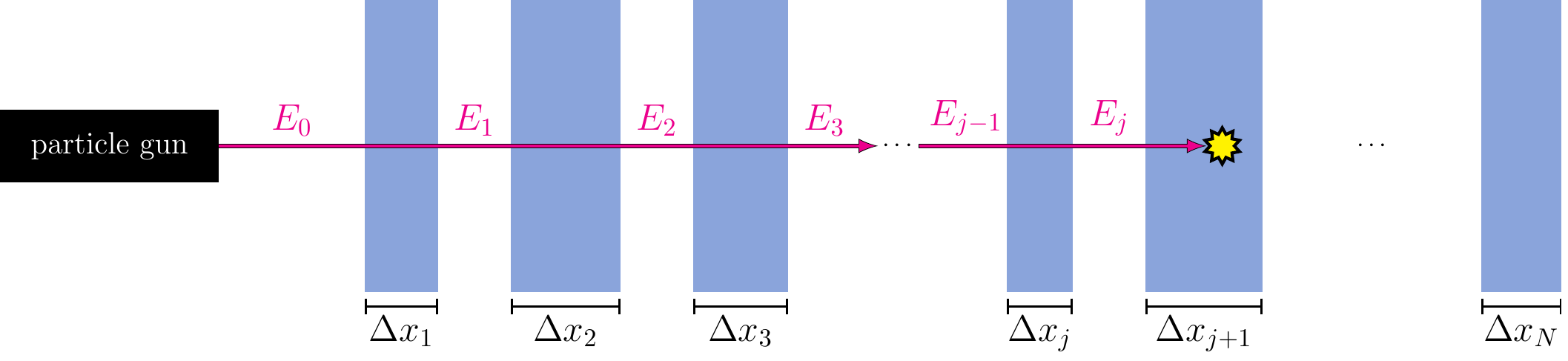}%
  \caption{A particle gun fires a single particle of initial kinetic energy
           $E_0$ at a series of targets.
           After propagating through the $i\textsuperscript{\,th}$ target,
           the particle is left with a kinetic energy of $E_i$;
           the particle deposits an energy of $E_i - E_{i-1}$ as it propagates
           through the $i\textsuperscript{\,th}$ target.
           The particle interacts in the
           $(j\!+\!1)\textsuperscript{\,th}$ target where it had a kinetic
           energy of $E_j$ before entering the $(j\!+\!1)\textsuperscript{\,th}$
           target.}%
  \label{fig:targets}%
\end{figure}%

This method of computing hadron--nucleus cross sections can be applied to data taken with a detector that is capable of keeping track of a particle's kinetic energy as it propagates through the target medium.  Fig.~\ref{fig:targets} depicts a cartoon of a particle gun firing a single particle of a known initial kinetic energy of $E_0$ at a series of targets.  The computation method described in Section~\ref{section:computation-method} can be applied to these targets, where instead of a particle gun shooting particles of fixed kinetic energies at single targets, we have a single particle passing through multiple targets.  As the particle propagates through each target, its kinetic energy can be determined by subtracting the energy deposited in the penetrated targets from its initial kinetic energy.  The \emph{non-interacting} kinetic energy of the particle at each target is taken to be its kinetic energy just before it enters into the target, given that there is no interaction; the \emph{interacting} kinetic energy of the particle is taken to be its kinetic energy just before it enters the target in which the interaction
occurs.  This can be done using an ensemble or beam of particles in order to minimize the statistical uncertainties in the hadron--nucleus cross section result.

For a segmented detector, this method is fairly straightforward and follows directly from Fig.~\ref{fig:targets}.  For a bubble chamber situated in a magnetic field, the momentum of particles can be determined by measuring the curvature of tracks; each track can be divided into small segments that are treated as a series of independent targets.  A homogeneous calorimeter with tracking capabilities, such as the TPC, can be divided into smaller targets based on the segment lengths of reconstructed tracks in the detector.  These smaller targets can then be treated as a series of independent targets as shown in Fig.~\ref{fig:targets}.  In the case of a TPC, the segment lengths of a reconstructed track is not necessarily equal to the uniform pitch of the wires and can vary slightly depending on the angle of the track segment relative to the plane of wires.  We conclude this section by noting that the mean target thickness or track segment length is ultimately determined by the properties of the detector and the reconstruction techniques used for the cross section measurement, and Fig.~\ref{fig:relative-error-vs-target-thickness} provides guidance on where the bias of the SR or WR methods becomes significant in an experiment.

\section{Conclusions}
The use of the LArTPC to study neutrino oscillations over short ($<\SI{1}{\km}$) and long ($>\SI{1000}{\km}$) baselines in modern neutrino experiments such as MicroBooNE, SBND, ICARUS, and DUNE has ushered in a demand for measurements of hadronic cross sections on argon in order to aid in the understanding of systematic uncertainties in LArTPC-based neutrino experiments.
The LArIAT and ProtoDUNE experiments each expose a fixed-target LArTPC to a controlled beam of charged particles for the purpose of measuring hadronic cross sections on argon.  The simple method of measuring cross sections on a thin slab of material requires an extension to be applied to a more voluminous target such as the LArTPC.

We have presented a method of computing hadron--nucleus cross sections that utilizes maximum likelihood estimation.  The observable data in this method are obtained by counting the number of particles that interact or do not interact, and measuring the particles' kinetic energies.  We have demonstrated that this method is robust even if the particle scattering data is taken from targets of different thicknesses or the target is a thick volume divided into slabs of various thicknesses.  The ability to robustly compute hadron--nucleus cross sections using maximum likelihood estimation has the potential to allow flexible design choices in fixed-target scattering experiments for measuring cross sections.




\section{Acknowledgments}
We would like to thank the LArIAT collaboration for the inspiration, and Anthony LaTorre for the fruitful discussions.
The computations in this paper were run on the FASRC Cannon cluster supported by the FAS Division of Science Research Computing Group at Harvard University.

\bibliographystyle{elsarticle-num}
\biboptions{sort&compress}
\bibliography{main}


\end{document}